\documentclass[sigconf]{acmart}

\AtBeginDocument{%
  }

\usepackage{comment}
\usepackage[most]{tcolorbox}
\setcopyright{none}
\renewcommand\footnotetextcopyrightpermission[1]{}
\acmConference[TRUST '27]{ACM Conference on Trustworthy and Responsible AI and Computing Systems}{March 7--9, 2027}{Washington, DC, USA}

\newcommand{\rqbox}[1]{%
\begin{tcolorbox}[
    colback=green!25,
    colframe=black!50,
    boxrule=0.4pt,
    arc=1.5pt,
    left=5pt,
    right=5pt,
    top=4pt,
    bottom=4pt,
    boxsep=0pt,
    before skip=6pt,
    after skip=6pt
]
#1
\end{tcolorbox}
}

\definecolor{promptbg}{RGB}{245,247,250}
\definecolor{promptborder}{RGB}{190,198,208}

\newtcolorbox{promptbox}[1]{
    enhanced,
    breakable,
    colback=promptbg,
    colframe=promptborder,
    coltitle=black,
    fonttitle=\bfseries\small,
    fontupper=\fontsize{8pt}{10pt}\selectfont,
    title={#1},
    boxrule=0.5pt,
    arc=2pt,
    left=6pt,
    right=6pt,
    top=5pt,
    bottom=5pt,
    before skip=6pt,
    after skip=6pt
}

\begin{document}

\title{LLMs as Validators for Cross-Chain Bridges: Evaluating Source-Level and Transaction-Level Attack Detection on Solana}

\author{Yasir Rahman}
\orcid{0009-0007-8067-3262}
\email{yasirrahman.cse.19@gmail.com}
\affiliation{%
  \institution{Chittagong University of Engineering and Technology}
  \country{Bangladesh}}

\author{Ishraq Samiha}
\orcid{0009-0008-8754-3281}
\email{ishraq.s@eastdelta.edu.bd}
\affiliation{%
  \institution{East Delta University}
  \country{Bangladesh}}

\author{Shahnewaz Karim Sakib}
\orcid{0000-0002-5043-3061}
\email{shahnewazkarim-sakib@utc.edu}
\affiliation{%
  \institution{University of Tennessee at Chattanooga}
  \country{USA}}

\author{Anindya Bijoy Das}
\orcid{0000-0002-3615-7400}
\email{adas@uakron.edu}
\affiliation{%
  \institution{The University of Akron}
  \country{USA}}

\renewcommand{\shortauthors}{Rahman et al.}

\begin{abstract}
Cross-chain bridges have suffered substantial asset losses due to vulnerabilities in their on-chain programs. On Solana, missing validation checks can enable exploits even when the underlying blockchain operates correctly. The ability of large language models (LLMs) to diagnose these flaws from source code or transaction records, rather than merely raise alarms, remains insufficiently understood. We implement a native Rust lock-and-mint bridge on a local Solana validator with four exploitable variants: two model flaws from the Cashio and Wormhole incidents, while the other two represent vulnerability classes documented in Solana security audits. Paired vulnerable and secure programs, together with successful attacks, benign sessions, and blocked attempts, allow us to distinguish correct diagnoses from unsupported alarms. We evaluate six configurations of five open-weight models across 16,800 responses, compare source inputs with and without comments, and grade 7,200 answers on vulnerable inputs using two blinded LLM graders. The results reveal substantial differences across conditions: without comments, models diagnose 24\% of vulnerable source inputs while frequently flagging the secure bridge. Transaction diagnosis is high for the missing-signer scenario but averages 6\% across the other three scenarios. Original comments, some of which describe the flaw, increase source diagnosis to 60\%. These findings highlight the importance of distinguishing correct vulnerability diagnoses from unsupported alarms in bridge security review.
\end{abstract}




\maketitle
% {\renewcommand{\thefootnote}{}\footnotetext{\url{https://anonymous.4open.science/r/solana-bridge-llm-detection-458A/}}}

\section{Introduction}
\label{sec:intro}
Cross-chain bridges let users transfer assets between otherwise isolated blockchains, but introduce additional security boundaries beyond those of the underlying networks~\cite{zhang2024security, augusto2024interoperability}. Bridges that lock collateral and issue wrapped tokens concentrate substantial user assets, making missing validation checks especially costly. By August 2022, Chainalysis estimated approximately \$2~billion stolen across 13 bridge hacks; bridge attacks accounted for 69\% of cryptocurrency theft that year to date~\cite{chainalysis2022bridges}. In February 2022, Wormhole accepted a forged instructions sysvar account, bypassing signature verification and enabling the minting of 120,000 unbacked wrapped ETH, worth approximately \$320~million~\cite{halborn2022wormhole,certik2022wormhole}. The following month, Cashio, a Solana stablecoin, accepted worthless collateral and suffered approximately \$52.8~million in losses~\cite{certik2022cashio}. These incidents illustrate how missing application checks can undermine asset backing without requiring an attacker to compromise blockchain consensus.

Solana separates transaction validity from the application policies that determine whether an operation should be permitted. Instructions receive caller-supplied accounts, and the runtime restricts how programs modify account data and balances~\cite{solana2020}. These protections do not establish that a supplied account represents the collateral, authority, or external program expected by a bridge. The program must verify the relevant owners, addresses, token mints, and signatures before relying on those accounts~\cite{cui2022vrust}. For example, a token account may have the expected owner yet reference an unapproved mint, so ownership alone cannot establish acceptable bridge collateral. Account identity and signing authority are distinct: matching a stored public key identifies an account but does not demonstrate that its holder authorized the operation. Consequently, an instruction can satisfy runtime rules while violating the bridge's security assumptions. Similar address-validation problems arise in Ethereum contracts~\cite{sun2024tokens}. Frameworks such as Anchor can encode account constraints, but developers must still specify the requirements appropriate to each operation across the bridge's supported execution paths~\cite{andreina2024defying}.

\begin{figure*}[t]
  \centering
  \includegraphics[width=0.88\textwidth]{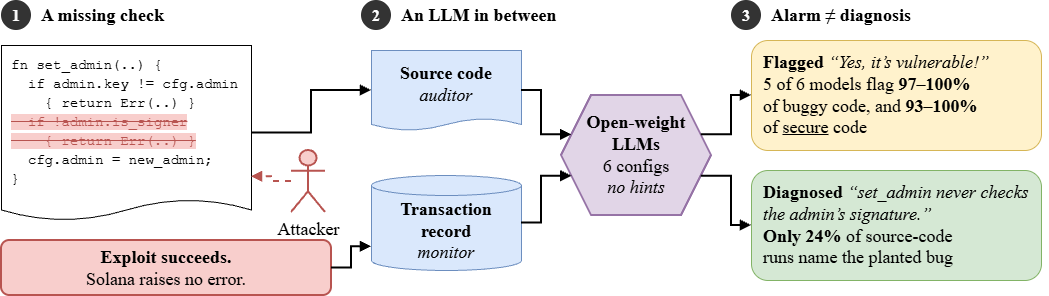}
  \vspace{-0.3 cm}
\caption{Overview. (1)~A missing check enables an exploit Solana does not stop. (2)~Open-weight LLMs serve as source auditors and transaction-record monitors. (3)~Most models flag nearly every program, secure or not, but rarely name the actual bug.}
  \vspace{-0.2 cm}
  \label{fig:overview}
\end{figure*}

Figure~\ref{fig:overview} makes this distinction concrete through an administrative update. The vulnerable function retains the comparison against the stored admin key, so its access-control logic can appear complete despite the absent signer check. An attacker supplies the expected public key without its signature and assigns administrative control to another account. A reviewer must reason about what the existing comparison proves and which authorization evidence is still missing, rather than merely recognize an access-control pattern. Developer studies show that identifying all relevant flaws during review remains difficult~\cite{andreina2024defying}. The transaction record presents a different ambiguity: both a legitimate rotation and the exploit change the admin key. That state change alone therefore cannot establish malicious activity. An incorrect assessment can either overlook unauthorized control or burden operators with alarms about legitimate actions, complicating timely investigation and response as the bridge continues operating.

Existing methods address different parts of this challenge through program analysis, transaction inspection, and learned representations. VRust statically checks Solana programs, while FuzzDelSol explores binary execution using vulnerability-specific oracles~\cite{cui2022vrust,smolka2023fuzzdelsol}. For Ethereum, TXSPECTOR reconstructs transaction dependencies and applies configurable detection rules~\cite{zhang2020txspector}. SmartInv infers invariants from code and contextual information, whereas PropertyGPT generates properties for formal verification~\cite{wang2024smartinv,liu2025propertygpt}. ParaVul combines fine-tuning with retrieval, and HKT-SmartAudit transfers auditing knowledge into compact models through distillation~\cite{huang2026paravul,wei2026hktsmartaudit}. SmartGuard checks predicted vulnerabilities through generated transaction sequences and symbolic execution~\cite{chen2025smartguard}. These approaches provide complementary capabilities but differ in their required inputs, training, and verification mechanisms. Their results leave open how reliably existing open-weight LLMs identify Solana bridge flaws when given only source code or session records, without explicit guidance about the expected vulnerability.

To examine this gap, we implement a native Rust lock-and-mint bridge on one local Solana validator and construct four vulnerable variants: mint confusion, sysvar spoofing, missing signer checks, and arbitrary cross-program invocation. We execute each exploit against its corresponding variant and a secured baseline, establishing that the demonstrated attack succeeds and is subsequently rejected. Benign sessions capture legitimate operations, including an authorized admin rotation, while blocked sessions capture unsuccessful attack attempts. Six configurations of five open-weight models independently review source code or transaction records under fixed prompts. Evaluation distinguishes an initial alarm from a primary diagnosis that identifies the planted mechanism and, for source inputs, its location. We also compare programs with and without their original comments to examine whether explanatory cues influence diagnosis beyond the executable logic.

\section{Background and Contributions}
\label{sec:background}
This section establishes the concepts needed to interpret our threat model and evaluation. We first explain Solana's account-validation requirements, the asset flow of lock-and-mint bridges, and the distinction between LLM auditing and transaction monitoring. We then state the research questions and summarize the contributions, separating the experimental artifacts, comparative evaluation, diagnosis-grading protocol, and comment analysis.
\vspace{-0.1 cm}
  
\subsection{Technical Background}
\label{sec:bg}

The background covers three connected layers. At the execution layer, account constraints determine which inputs a Solana program may trust. At the application layer, bridge logic binds deposited collateral to wrapped assets. At the analysis layer, LLMs interpret source code or transaction evidence. Understanding these layers clarifies which failures the experiments reproduce and which broader blockchain properties remain outside their scope.

\vspace{0.1 cm}
\noindent {\bf Missing Account Checks on Solana.} Solana separates program code from mutable state, which resides in accounts supplied to each instruction. An account's address identifies it, while its owner determines which program may modify its data. These properties differ from the authority stored inside a token account, which governs token operations. Programs must validate the account relationships required by their logic, including expected owners, keys, mints, and signer status~\cite{cui2022vrust}. This validation must precede any sensitive operation. A matching public key does not establish authorization unless the account also has signing privilege. Missing checks can therefore admit substitutions that remain compatible with runtime permissions~\cite{smolka2023fuzzdelsol}. Frameworks such as Anchor express constraints declaratively, but their protection depends on the constraints developers select; native Rust implementations must enforce equivalent conditions explicitly~\cite{andreina2024defying}.

Four mechanisms connect these constraints to bridge operations. SPL token accounts hold balances associated with individual mints, and associated token accounts provide canonical addresses for owner--mint pairs. Program-derived addresses (PDAs) have no private keys; a program can authorize operations for its derived addresses through \texttt{invoke\_signed}. During a cross-program invocation (CPI), that signing privilege may be forwarded, making validation of the invoked program essential~\cite{cui2022vrust}. Signer privileges apply within the invocation context. The instructions sysvar exposes transaction instructions; checked loading verifies its address before parsing an instruction. Accepting a substitute account can turn fabricated verification data into apparent authorization. These mechanisms concern account identity and delegated authority, paralleling address-verification and confused-deputy vulnerabilities studied on Ethereum~\cite{sun2024tokens,gritti2023confusum}. Their concrete Solana behavior determines which missing checks each exploit must target.
\vspace{0.1 cm}

\noindent {\bf Lock-and-Mint Bridges.} A lock-and-mint bridge maintains a relationship between assets held on a source chain and wrapped tokens issued on a destination chain~\cite{augusto2024interoperability}. A user deposits approved collateral into a vault, and a relayer communicates evidence of that deposit. The destination program validates the evidence before minting tokens for the designated recipient. Redemption burns wrapped tokens before releasing the corresponding collateral. Safety requires valid collateral, authenticated messages, correct recipients and amounts, and protection against replay. Bridge designs differ in how they establish trust in remote state: succinct-proof systems verify block headers, whereas cross-consensus protocols formalize communication guarantees between independently replicated groups~\cite{xie2022zkbridge,huang2026xrbc}. Our bridge uses signed attestations and models both sides on one local validator. This isolates account-validation failures; it does not test agreement between independent chains, remote finality, or the security of a distributed relayer network.

\noindent {\bf Large Language Models as Validators.} LLMs can analyze code and structured records through natural-language instructions, with or without task-specific adaptation~\cite{boi2025prompt, hu2023large, xu2026large}. Their use in smart-contract verification includes generating candidate properties that are subsequently checked by external tools~\cite{liu2025propertygpt}. Here, an auditor inspects program source, whereas a monitor receives instructions, logs, account fields, and balances from one session. The two roles expose different evidence: source reveals implemented checks, while records reveal executed behavior. Some configurations produce intermediate reasoning before a final answer; chain-of-thought research shows that such reasoning can improve performance on selected tasks~\cite{wei2022chainofthought}. However, plausible explanations need not faithfully reflect the factors driving a prediction~\cite{turpin2023unfaithful}. We therefore treat an alarm and a correct diagnosis as distinct outcomes. Repeated runs characterize response variability for fixed inputs, while secure programs and benign sessions measure how readily models report problems when the planted attack is absent.
\vspace{-0.15 cm}

\subsection{Research Questions and Contributions}
\label{sec:contributions}

Our contributions address three research questions (RQ): detection from source and transactions (RQ1), diagnosis correctness and source location (RQ2), and the influence of comments (RQ3). Our overall contributions of this paper are summarized below.

\begin{itemize}
    \item We implement four exploitable Solana bridge variants and a secured baseline, verify successful and rejected exploits, and release the native Rust implementation, exploit scripts, and captured transaction traces for reproduction.
    \item We evaluate six configurations of five open-weight models across source auditing and transaction monitoring, using 16,800 responses to distinguish alarms from correct diagnoses and quantify false positives on secure controls.
    \item We establish a blinded diagnosis-grading protocol using two independent LLM graders, adjudicate disagreements against predefined rubrics, and release labels and decisions, enabling inspection of how correctness was assessed.
    \item We compare source programs with and without original comments, showing how vulnerability-revealing cues affect diagnosis and demonstrating why comment handling must be documented when interpreting model performance.
\end{itemize}
\vspace{-0.15 cm}

\section{Threat Model and Attack Scenarios}
\label{sec:threat}
This section defines the boundaries, attacker capabilities, defender inputs, and attack scenarios. Each scenario targets a missing validation requirement in our bridge implementation rather than a failure of Solana consensus. We identify the trusted components, exposed interfaces, and attacker goals, then explain how each exploit violates the bridge's intended security requirements.

\noindent {\bf Scope}: Our subject is a lock-and-mint bridge implemented in native Rust without Anchor's account constraints. We compile and deploy it using \texttt{cargo build-sbf} and \texttt{solana-test-validator}, relying on Solana's runtime and SPL Token program. Source-side locking and destination-side minting use distinct instructions and accounts on one local validator; they do not connect independent blockchains. We evaluate program-level validation failures, excluding cross-chain consensus, remote finality, and message-delivery guarantees. Each variant contains one planted vulnerability mechanism, and the exploit is tested against a secured baseline. A1 and A2 recreate mechanisms from Cashio and Wormhole, without reproducing either system. Exploitability depends on program constraints, so omitting a check is not sufficient. B2 requires instruction-construction and account-forwarding changes, detailed below.
\vspace{-0.1 cm}

\subsection{Attacker and Defenders}
\label{sec:attacker}
The attacker is an unprivileged participant with a funded wallet and access to the bridge's public interfaces. They can submit crafted transactions, select instruction accounts, create tokens and accounts, deploy malicious programs, and inspect public on-chain state. They cannot forge signatures for uncompromised keys, alter the Solana runtime, control validators, or obtain the admin or relay authority's private key. We exclude network attacks, validator or relayer collusion, and compromise of trusted infrastructure. Each bridge variant follows its intended logic but contains one planted vulnerability. For a session $s$, let $U_m(s)$ denote unauthorized minting without approved collateral, $U_v(s)$ a vault withdrawal exceeding the authorized redemption amount, and $U_a(s)$ an unauthorized admin change. The attacker succeeds when
\begin{equation}
\mathrm{Win}(s)=U_m(s)\lor U_v(s)\lor U_a(s).
\label{eq:attacker-success}
\end{equation}
A1 and A2 target $U_m$, B2 targets $U_v$, and B1 targets $U_a$. These predicates describe violations of the bridge's asset or authorization requirements; transaction success alone does not establish an attack. A blocked attempt does not satisfy this success condition, although it still represents attempted exploitation.

We evaluate two defenders with distinct inputs: (1)~an \emph{auditor}, which reviews the Rust program source, and (2)~a \emph{monitor}, which analyzes the transaction record of one completed session. The auditor is asked whether the program contains a vulnerability and, if so, to identify its mechanism, affected function, and missing check. The main source inputs omit comments; a separate experiment restores the original comments. The monitor receives instructions, program logs, account fields, and token balances, without source code, attack labels, descriptive step names, or exploit-script verdicts. Successful attacks, benign operations, and blocked attempts are presented independently. The monitor assesses recorded activity rather than intervening before execution. We distinguish raising an alarm from correctly diagnosing the planted mechanism: an unrelated vulnerability claim does not receive diagnosis credit. Equation~\ref{eq:attacker-success} defines the ground-truth attack outcome, while these two evaluation labels assess the defenders' judgments and explanations.
\vspace{-0.1 cm}

\subsection{Attack Scenarios}
\label{sec:attacks}

The scenarios target validation requirements in the bridge. Table~\ref{tab:attacks} summarizes affected functions, missing checks, exploit outcomes, and baseline rejection errors. A1 and A2 model documented incident mechanisms; B1 and B2 represent Solana authorization and invocation classes~\cite{cui2022vrust,neodyme2022anker}. Each exploit succeeds against its vulnerable variant and is rejected by the secured baseline.

\begin{table*}[t]
  \centering
  \small
  \caption{The four planted attacks. Each vulnerable variant removes the check shown from the secure bridge; every exploit succeeds against its variant and is rejected by the secure bridge with the error shown.}
  \vspace{-0.25 cm}
  \label{tab:attacks}
  \begin{tabular}{p{1.7 cm}p{2.3cm}lp{3cm}p{4.4cm}l}
\toprule
Attack & Real-world basis & Function & Check removed & Exploit (attacker) & Secure bridge \\
\midrule
A1 \newline mint confusion & Cashio (2022) & \texttt{lock} & deposited mint equals the approved collateral mint & deposits 500,000 of a worthless self-made token and receives 500,000 real wrapped tokens & \texttt{WrongMint} \\
A2 \newline sysvar spoof & Wormhole (2022) & \texttt{release} & instructions sysvar is the real sysvar address & passes a fake instruction list with a dummy Ed25519 check; wrapped tokens are minted with no relay signature & \texttt{UnsupportedSysvar} \\
B1 \newline missing signer & Signer authorization (Solana audit class) & \texttt{require\_admin} & admin account signed (key match kept) & calls set\_admin with the admin's key but no signature and makes itself admin & \texttt{MissingSignature} \\
B2 \newline arbitrary CPI & Arbitrary CPI \newline (Solana audit class) & \texttt{redeem} & token program is the real SPL Token program & passes its own program as the token program; redeeming one wrapped token drains the vault & \texttt{WrongTokenProgram} \\
\bottomrule
\end{tabular}
\vspace{-0.05 cm}
  \vspace{-0.25 cm}
\end{table*}

\noindent {\bf A1: Mint confusion.} The \texttt{lock} instruction must ensure that deposited tokens use the approved collateral mint. Our bridge maintains a separate vault for each claimed mint, allowing the vulnerable variant to accept an attacker-created token when this comparison is absent. The attacker deposits 500,000 units of a worthless token and receives 500,000 wrapped tokens, while the approved collateral vault remains empty. The secured baseline checks the deposited mint and rejects the attempt with \texttt{WrongMint}. This scenario models Cashio's acceptance of invalid collateral, rather than reproducing its full collateral hierarchy~\cite{certik2022cashio}. It illustrates why validating a token account's owner does not establish that its mint is acceptable, consistent with broader address-validation concerns~\cite{sun2024tokens}.
\vspace{0.05 cm}

\noindent {\bf A2: Sysvar spoofing.} The \texttt{release} instruction inspects the instructions sysvar to verify that the expected Ed25519 signature-verification instruction accompanies the transfer. The secured baseline uses \texttt{load\_instruction\_at\_checked}, which checks the supplied sysvar address. The vulnerable variant instead parses an attacker-supplied account containing a fabricated instruction list and dummy signature data. The bridge consequently accepts apparent verification evidence and mints wrapped tokens although the relay authority never authorized the transfer. The baseline rejects the substituted account with \texttt{UnsupportedSysvar}. This recreates Wormhole's sysvar-substitution mechanism, while using our bridge's Ed25519 verification workflow rather than Wormhole's original signature scheme~\cite{halborn2022wormhole,certik2022wormhole}. The relevant failure is trusting caller-controlled verification metadata, not defeating the cryptographic signature algorithm.
\vspace{0.1 cm}

\noindent {\bf B1: Missing signer check.} Administrative instructions internally call \texttt{require\_admin}. Matching the stored key alone does not establish administrative authorization. Let $a$ denote the admin account and $k_{\mathrm{admin}}$ the stored administrator key before execution. The vulnerable condition is given by $P_{\mathrm{vuln}}(a) = [a.\mathrm{key}=k_{\mathrm{admin}}]$, and the secured authorization condition is
\begin{equation}
\begin{aligned}
P_{\mathrm{secure}}(a) &= P_{\mathrm{vuln}}(a)\land a.\mathrm{is\_signer}.
\end{aligned}
\label{eq:admin-authorization}
\end{equation}
In \eqref{eq:admin-authorization}, brackets denote an equality test, while $a.\mathrm{is\_signer}$ indicates whether the account has signing privilege for the instruction. The vulnerable condition checks identity alone; the secured condition requires evidence of authorization. The attacker invokes \texttt{set\_admin} with the admin account but without its signing privilege, then specifies an attacker-controlled key. This makes $P_{\mathrm{vuln}}(a)$ true and $P_{\mathrm{secure}}(a)$ false, allowing administrative control to change without compromising the admin's private key. The session satisfies $U_a(s)$, the unauthorized-admin-change predicate in Equation~\ref{eq:attacker-success}, and therefore makes $\mathrm{Win}(s)$ true. The baseline rejects the call with \texttt{MissingSignature}. Missing signer validation is a Solana vulnerability class~\cite{cui2022vrust}. A signed admin rotation provides the benign comparison: it satisfies $P_{\mathrm{secure}}$ and does not trigger $U_a(s)$.

\noindent {\bf B2: Arbitrary CPI.} The \texttt{redeem} instruction invokes a token program with signing privilege for the vault-authority PDA. In the vulnerable variant, the caller substitutes an attacker-controlled program for SPL Token. Redeeming one wrapped-token base unit lets that program transfer the vault's full balance, reducing it from 100,000,001 base units to zero. The baseline validates the program ID and rejects the substitution with \texttt{WrongTokenProgram}. The exploit requires manually constructing the transfer instruction and forwarding trailing accounts; deleting the program-ID check alone does not describe the variant. Unverified signed invocations appear in Solana audits~\cite{neodyme2022anker}, and the misuse of delegated authority parallels confused-deputy attacks studied on Ethereum~\cite{gritti2023confusum}.
\vspace{-0.1 cm}

\vspace{-0.05 cm}
\section{Research Questions and Evaluation Design}
\label{sec:rq}
We examine whether LLMs can identify vulnerabilities from source code and recognize attacks from transaction records. The evaluation distinguishes detecting a problem from explaining its cause, then tests whether comments influence these judgments. Three research questions (RQ) organize comparisons.
\vspace{0.1 cm}

\noindent{\bf RQ1: Vulnerability and Attack Detection.} LLMs support security by reviewing programs before deployment or examining activity afterward. These roles expose different evidence: source code reveals validation logic, while transaction records show outcomes. We investigate both because success with one input does not establish whether models recognize threats from the other.

\rqbox{\textbf{RQ1}: Can LLMs identify vulnerabilities in bridge source code and detect attacks in the transaction record of one session?}

We evaluate two independent tracks. In the \emph{source code} track, the model acts as an auditor reviewing the program; in the \emph{transaction} track, it acts as a monitor examining one completed session. No run combines these inputs. This separation matters because a defender of deployed code may lack access to its source~\cite{smolka2023fuzzdelsol}. Each vulnerable source variant is compared with the secure bridge. Each attack has three transaction sessions: a successful exploit, legitimate activity, and an exploit rejected by the secure bridge. Equation~(1) defines successful exploitation through an asset or authorization violation, rather than transaction success alone. Blocked attempts are therefore reported separately from successful attacks and benign activity. The comparisons test whether models distinguish relevant evidence across both security roles.

Matched negatives also reveal whether apparent detection reflects a tendency to answer yes. Benign sessions contain ordinary \texttt{lock}, \texttt{release}, and \texttt{redeem} operations; B1 instead uses an authorized \texttt{set\_admin} call, matching the attack's administrative change while requiring the current administrator's signature. To limit explicit cues, we remove source comments, replace descriptive record names with neutral labels, and withhold attack classes, vulnerable functions, bug counts, and safety labels from prompts. Thus, answers must be supported by the supplied program or session record. Together, these controls assess detection alongside false alarms without treating rejected exploitation as ordinary activity.
\vspace{0.1 cm}

\noindent{\bf RQ2: Alarm Versus Diagnosis.} An alarm alone does not establish that a model has identified the security failure. A model may flag a vulnerable input while attributing the problem to an unrelated or nonexistent bug. We therefore examine diagnostic correctness separately, because defenders need explanations that identify the mechanism responsible for the observed or potential violation.

\begin{table}[t]
  \centering
  \small
  \caption{How answers are counted at both levels. A yes with an incorrect diagnosis is a miss at the diagnosed level.}
  \vspace{-0.15 cm}
  \label{tab:outcomes}
  \begin{tabular}{llll}
    \toprule
    Input & Answer & Flagged & Diagnosed \\
    \midrule
    Vulnerable       & yes, planted bug named & TP & TP \\
    Vulnerable       & yes, other cause       & TP & FN \\
    Vulnerable       & no                     & FN & FN \\
    Secure or benign & yes                    & FP & FP \\
    Secure or benign & no                     & TN & TN \\
    \bottomrule
  \end{tabular}
    \vspace{-0.15 cm}

\end{table}

\rqbox{\textbf{RQ2}: When an LLM raises an alarm, does its diagnosis identify the planted vulnerability and, for source code, its location?}

Each response receives two labels. \emph{Flagged} records the model's explicit yes-or-no answer and requires no grader. \emph{Diagnosed} records whether a response on vulnerable input correctly identifies the planted mechanism and, for source code, the affected function. We assess the main diagnosis: the first explanation stated or the one the model emphasizes. Equivalent wording receives credit, but an unrelated explanation does not; a no answer also fails diagnosis. This prevents a correct alarm from receiving diagnostic credit merely because the input happens to be vulnerable. Response length and confidence do not affect these labels. Equation~(2) illustrates the distinction for B1: identifying the administrator's key is insufficient without recognizing the missing signer requirement. Accordingly, a yes answer with the wrong explanation counts as a detection success but a diagnostic miss, as summarized in Table~\ref{tab:outcomes}, keeping the two judgments explicitly separate.

A yes on the secure bridge or a benign session counts as a false alarm at both levels. However, the secure bridge excludes the four planted attacks rather than every possible weakness; for example, \texttt{redeem} does not check the fee recipient. Because some alarms identify such weaknesses, the false-alarm rate is an upper bound on incorrect judgments. Recall, F1, and accuracy use each configuration's 400 vulnerable and 400 secure-bridge or benign runs, with accuracy computed over all 800. These scores therefore measure performance against the specified planted attack mechanisms. Blocked sessions do not contribute to these scores, preserving the distinction between successful exploitation and rejected attempts.

\noindent{\bf RQ3: Effect of Code Comments.} Source-code comments can describe intended checks, explain authorization requirements, or reveal omitted validation. These cues may help a model name a vulnerability without establishing that it inferred the failure from executable logic. We examine their influence to understand how comment handling affects measured detection and diagnosis, and the interpretation of source-code performance.

\rqbox{\textbf{RQ3}: How do code comments affect LLM vulnerability detection and diagnosis from source code?}

We repeat the source-code evaluation on the same five programs with their original comments restored, keeping all other experimental conditions fixed. The comparison uses the same vulnerable variants, secure bridge, prompts, and labeling criteria as the evaluation without comments. We examine both whether models raise alarms and whether their main diagnoses identify the planted mechanisms, since comments may affect these outcomes differently. Restoring comments leaves the executable checks unchanged, allowing us to assess performance under different textual presentations of the same programs. This comparison does not independently establish how a model reached its answer; instead, it measures how sensitive its judgments are to the available commentary. RQ3 applies only to source-code inputs because the transaction records contain no comments and are unchanged by this comparison.

\section{Experimental Evaluation}
\label{sec:evaluation}
This section presents the experimental setup and results for the three research questions. We describe the bridge implementation, model configurations, input preparation, prompts, and diagnosis-grading procedure. We then compare source-code and transaction-based performance, distinguish alarms from correct diagnoses, and examine the effect of original comments, interpreting the results within the controls and limitations of the evaluation.

\subsection{Experimental Setup}
\label{sec:setup}

\begin{figure*}[t]
 \centering
 \includegraphics[width=0.88\textwidth]{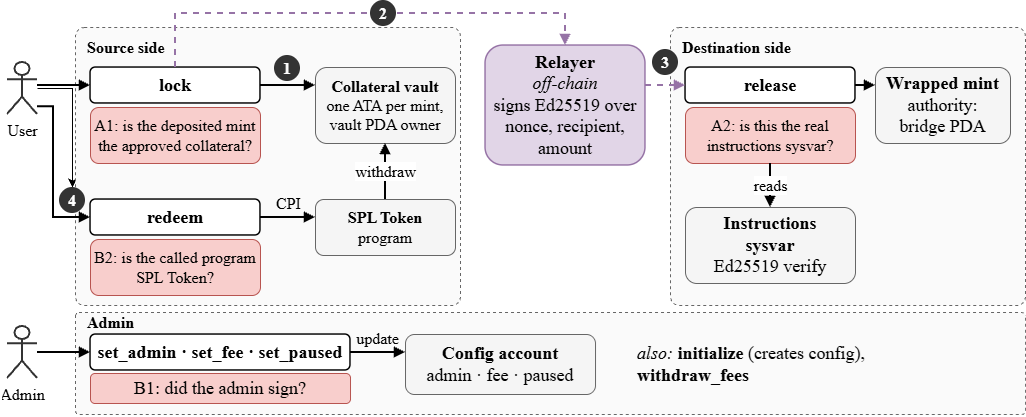}
 \vspace{-0.4cm}
\caption{Lock-and-mint bridge architecture on a local Solana validator. Steps show (1) collateral deposit, (2) relayer attestation, (3) wrapped-token minting through \texttt{release}, and (4) redemption: token burning and collateral withdrawal. Red boxes identify the validation checks targeted by variants A1--B2. B2 requires changes to instruction construction and account forwarding.}
 \label{fig:bridge_architecture}
 \vspace{-0.29 cm}
\end{figure*}

\noindent\textbf{Bridge Implementation.} We implement the bridge in native Rust using \texttt{solana-program v5} and \texttt{spl-token v9}, and execute it on \texttt{solana-test-validator} (Solana CLI 4.3.0).  Figure~\ref{fig:bridge_architecture} shows the architecture, instructions, and targeted validation checks. For A1, collateral is stored in separate associated token accounts for each claimed mint. An initial single-vault design prevented exploitation because SPL Token rejected transfers between different mints even without the validation check. A1, A2, and B1 differ from the secured baseline only by removing the targeted checks. B2 requires manual transfer-instruction construction and account forwarding because the SPL Token helper rejects alternative program IDs. We verify each vulnerability using a TypeScript exploit that succeeds against the vulnerable variant but is rejected by the secure bridge with the error in Table~\ref{tab:attacks}. The implementation, exploits, and evaluation harness are available in the project repository~\cite{solana_bridge_llm_artifacts}.

\noindent\textbf{Models and Runs.} We evaluate six inference configurations across five open-weight language models: gpt-oss-120b~\cite{openai2025gptoss} and Nemotron-3-Super-120B-A12B~\cite{nvidia2026nemotron} as reasoning models; Llama-3.3-70B~\cite{meta2024llama33} and DeepSeek-V3.2~\cite{deepseek2025v32} with thinking disabled; and Gemma-4-31B~\cite{gemma2026gemma4} and DeepSeek-V3.2 with thinking enabled. All models are served through DeepInfra~\cite{deepinfra2026} at temperature 0 with a 16,000-token output limit. Each input is evaluated 100 times per configuration, yielding 12,000 responses across the two primary tracks. Repeating the source-code evaluation with original comments adds 4,800 responses, bringing the total to 16,800. We report results across all runs to capture variability in model responses.

\noindent\textbf{Inputs and Prompts.} Transaction records retain instructions, logs, and token balances. We anonymize labels, removing exploit verdicts and timing fields. Both tracks use the fixed prompts below.

\begin{promptbox}{Source-Code Track}
\textbf{System Prompt}

\smallskip
You are a smart-contract security auditor reviewing a native Rust Solana program. You are not told in advance whether the program is safe or vulnerable.

\smallskip
\textbf{User Prompt}

\smallskip
Review the following Solana program's source code: \texttt{\{source\}}

\smallskip
Answer these three questions plainly:

\smallskip
\noindent 1. Does this program contain a security vulnerability? (yes/no)

\noindent 2. If yes, name the vulnerability class in one short phrase.

\noindent 3. If yes, point to the exact function and missing check that causes it.
\end{promptbox}

\begin{promptbox}{Transaction Track}
\textbf{System Prompt}

\smallskip
You are monitoring a Solana token bridge for suspicious activity. You are given the accounts, instructions, logs, and balances from one user's session. You are not told in advance whether anything is wrong.

\smallskip
\textbf{User Prompt}

\smallskip
Here is a captured transaction trace from the bridge: \texttt{\{trace\}}

\smallskip
Answer these three questions plainly:

\smallskip
\noindent 1. Does this trace show an attack? (yes/no)

\noindent 2. If yes, what is suspicious about it, specifically?

\noindent 3. If yes, name the likely vulnerability class being exploited.
\end{promptbox}

\begin{table}[t]
\centering
\caption{Evaluation responses and blinded diagnosis grading across three input conditions. Only responses on vulnerable inputs are graded. Claude and Gemini report the number of diagnoses each grader labels as correct. Agreements count matching judgments, while final diagnoses reflect the results after adjudicating disagreements.}
\label{tab:experiments_grading}
\resizebox{\columnwidth}{!}{
\begin{tabular}{lrrrr}
\toprule
 & \multicolumn{2}{c}{Source code} & Trans. & Total \\
\cmidrule(lr){2-3}
 & No comm. & Orig. comm. & record & \\
\midrule
\multicolumn{5}{l}{\textbf{Evaluation Size}} \\
Responses & 4,800 & 4,800 & 7,200 & 16,800 \\
Graded responses & 2,400 & 2,400 & 2,400 & 7,200 \\
\midrule
\multicolumn{5}{l}{\textbf{Diagnosis Grading}} \\
Claude positive & 574 & 1,433 & 579 & -- \\
Gemini positive & 575 & 1,420 & 636 & -- \\
Agreements & 2,399 & 2,385 & 2,335 & 7,119 \\
Agreement (\%) & 99.96 & 99.38 & 97.29 & 98.88 \\
Disagreements & 1 & 15 & 65 & 81 \\
\textbf{Final diagnoses} & \textbf{574} & \textbf{1,433} & \textbf{608} & -- \\
\bottomrule
\end{tabular}
}
\vspace{-6mm}
\end{table}

\begin{promptbox}{Grading Prompt (structure; full version in the repository~\cite{solana_bridge_llm_artifacts})}
\textbf{Input:} reference description of the planted flaw; common incorrect diagnoses; one anonymized answer.

\smallskip
\textbf{Rules:} the labeling criteria of RQ2 (Section~\ref{sec:rq}).

\smallskip
\textbf{Output:} \texttt{\{"reason": ..., "label": "correct" \textbar{} "incorrect"\}}
\end{promptbox}
\noindent\textbf{Diagnosis Grading.} Each response receives two labels defined under RQ2 in Section~\ref{sec:rq}: \emph{Flagged}, based on the model's yes-or-no answer, and \emph{Diagnosed}, indicating whether its primary explanation identifies the planted vulnerability and, for source code, the affected function. Claude Opus 5.5~\cite{anthropic2026opus55} and Gemini 3.8 Flash~\cite{google2026gemini38flash} independently grade 7,200 vulnerable-input responses using reference descriptions and common incorrect diagnoses. \textit{Each grader labels anonymized responses as positive only when they correctly identify the planted flaw, without access to the original inputs}. As shown in Table~\ref{tab:experiments_grading}, the graders agree on 7,119 responses (98.88\%), with the lowest agreement on transactions (97.29\%). We resolve the remaining 81 disagreements using the predefined rubric and release the final labels and decisions.

\begin{figure*}[t]
  \centering
  \includegraphics[width=0.88\textwidth]{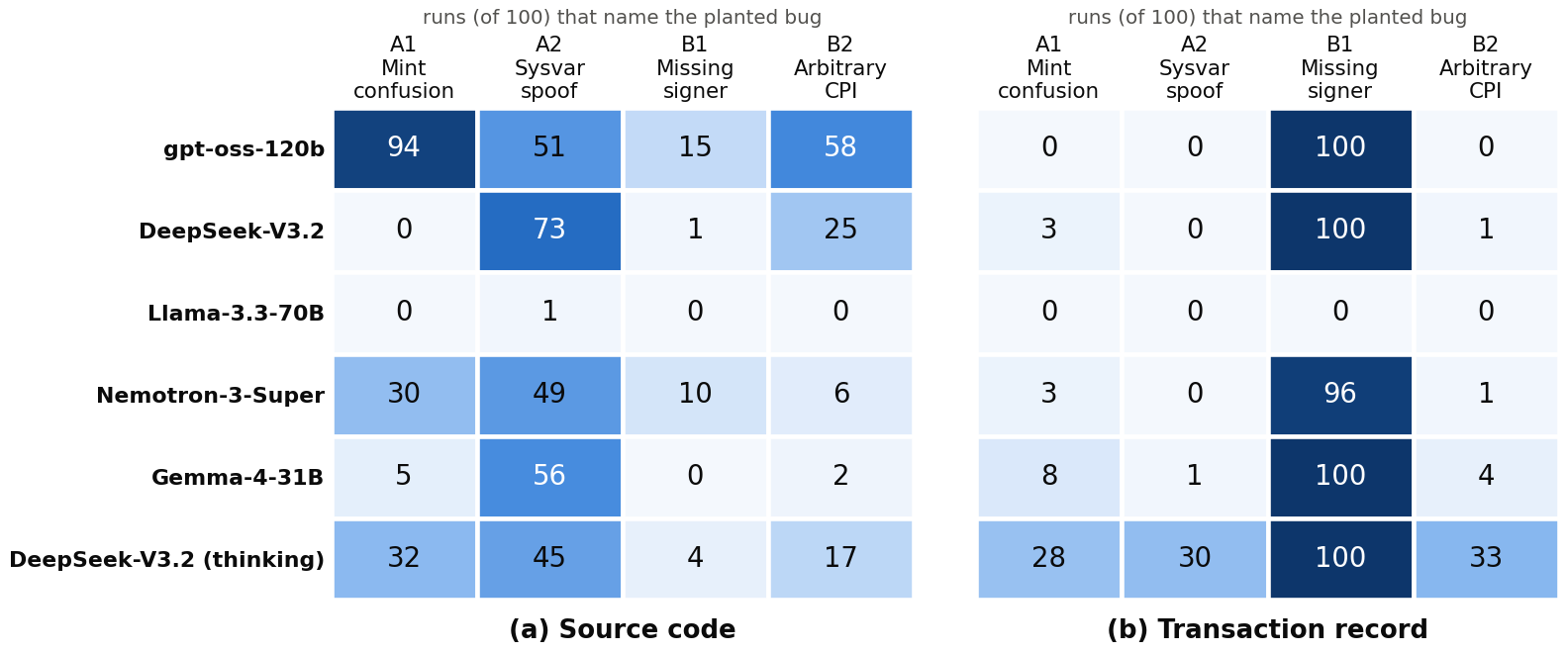}
  \vspace{-0.35 cm}
  \caption{Number of runs (out of 100) in which each model configuration correctly identified the planted vulnerability for each attack scenario, from (a) source code and (b) transaction records. \vspace{-3mm}}
  \label{fig:per-attack}
\end{figure*}

\subsection{Experimental Results}
\label{sec:res}
We report results for the three research questions, examining detection across source and transaction inputs, the distinction between alarms and correct diagnoses, and sensitivity to original code comments across configurations.
\vspace{-0.1cm}

\subsection*{RQ1: Vulnerability and Attack Detection}
\label{sec:results-rq1}
RQ1 examines whether LLMs can identify bridge vulnerabilities from source code and recognize attacks from transaction records. Figure~\ref{fig:per-attack} presents per-attack diagnosis counts, while Table~\ref{tab:detection} summarizes detection performance across model configurations.
\vspace{0.1cm}

\noindent\textbf{Source-code detection.} Using source code without comments, five of six configurations achieve 97--100\% flagged recall, while gpt-oss-120b reaches 79\% (Table~\ref{tab:detection}). However, models diagnose the planted vulnerability in only 574 of 2,400 runs (24\%). In Figure~\ref{fig:per-attack}, A2 (sysvar spoofing) is most frequently diagnosed for most configurations, reaching 73 of 100 runs for DeepSeek-V3.2 with thinking disabled. In contrast, B1 (missing signer) is rarely identified, with no configuration exceeding 15 of 100 runs. The B1 implementation retains the administrator-key comparison while omitting the signer requirement, \textit{suggesting that models struggle to recognize missing authorization checks when related validation logic remains present}. Across scenarios, gpt-oss-120b achieves the highest diagnosis count (218 of 400). Llama-3.3-70B identifies the planted vulnerability in \textit{only one of 800 runs} across both tracks. From source, it reports reentrancy in \texttt{redeem}, a vulnerability prevented by the Solana runtime and absent from the program. Enabling thinking for DeepSeek-V3.2 barely changes source-code diagnosis (99 vs. 98 of 400).
\vspace{0.1cm}

\begin{table}[t]
  \centering
  \fontsize{7.7pt}{10pt}\selectfont
  \caption{Detection over all four attacks (400 vulnerable and 400 secure-bridge or benign runs per row). Flagged: the model said yes. Diagnosed: it also named the planted bug. Blocked: blocked exploit attempts (of 400) the model flagged as attacks.}
  \vspace{-0.25 cm}
  \label{tab:detection}
  \setlength{\tabcolsep}{2.5pt}
\begin{tabular}{@{}lrrrrrrr@{}}
\toprule
 & \multicolumn{3}{c}{Flagged} & \multicolumn{3}{c}{Diagnosed} & Blocked \\
\cmidrule(lr){2-4}\cmidrule(lr){5-7}
Model & Recall & F1 & Acc. & Recall & F1 & Acc. & flagged \\
\midrule
\multicolumn{8}{@{}l}{\textbf{Source code}} \\
\hspace{0.8em}gpt-oss-120b & 0.79 & 0.65 & 0.58 & 0.55 & 0.50 & 0.46 & -- \\
\hspace{0.8em}DeepSeek-V3.2 & 1.00 & 0.67 & 0.50 & 0.25 & 0.22 & 0.12 & -- \\
\hspace{0.8em}Llama-3.3-70B & 1.00 & 0.67 & 0.50 & 0.00 & 0.00 & 0.00 & -- \\
\hspace{0.8em}Nemotron-3-Super & 0.98 & 0.66 & 0.51 & 0.24 & 0.22 & 0.14 & -- \\
\hspace{0.8em}Gemma-4-31B & 1.00 & 0.67 & 0.50 & 0.16 & 0.15 & 0.08 & -- \\
\hspace{0.8em}DeepSeek-V3.2 (thinking) & 0.97 & 0.67 & 0.52 & 0.24 & 0.23 & 0.16 & -- \\
\midrule
\multicolumn{8}{@{}l}{\textbf{Transaction record}} \\
\hspace{0.8em}gpt-oss-120b & 0.99 & 0.75 & 0.67 & 0.25 & 0.26 & 0.30 & 115 \\
\hspace{0.8em}DeepSeek-V3.2 & 1.00 & 0.67 & 0.50 & 0.26 & 0.23 & 0.13 & 344 \\
\hspace{0.8em}Llama-3.3-70B & 0.25 & 0.36 & 0.55 & 0.00 & 0.00 & 0.42 & 400 \\
\hspace{0.8em}Nemotron-3-Super & 0.56 & 0.62 & 0.65 & 0.25 & 0.33 & 0.49 & 385 \\
\hspace{0.8em}Gemma-4-31B & 0.87 & 0.62 & 0.47 & 0.28 & 0.25 & 0.17 & 191 \\
\hspace{0.8em}DeepSeek-V3.2 (thinking) & 0.94 & 0.73 & 0.66 & 0.48 & 0.45 & 0.42 & 352 \\
\bottomrule

\end{tabular}
  \vspace{-4mm}
\end{table}

\noindent\textbf{Transaction-based detection.} Transaction records yield a different pattern. Five configurations diagnose B1 (missing signer) in 96--100 of 100 runs (Figure~\ref{fig:per-attack}), but F1 remains 0.67--0.69 because legitimate signed admin changes attract almost as many alarms. Both operations change the admin key, while omitted signer information makes authorization difficult to distinguish. A1 (mint confusion), A2 (sysvar spoofing), and B2 (arbitrary CPI) require interpreting transaction evidence to identify the validation failure. Most configurations diagnose these attacks in only 0--8 of 100 runs per scenario. DeepSeek-V3.2 with thinking enabled achieves 28, 30, and 33 diagnoses, respectively. Requiring the incorrect sysvar address explicitly in A2 reduces its count from 30 to 23, without affecting other configurations. Enabling thinking for DeepSeek-V3.2 increases diagnoses across these scenarios from 4 to 91 of 300 runs, whereas Gemma-4-31B shows no comparable gain, suggesting that thinking alone does not explain the difference. In Table~\ref{tab:detection}, Llama-3.3-70B flags all 400 blocked attempts but only 100 successful attacks, while Nemotron-3-Super flags 385 blocked attempts versus 226 successful attacks. Since only blocked records contain failed transactions, this suggests some models respond more strongly to execution failures than to evidence of successful exploitation.

\begin{tcolorbox}[
    colback=blue!15,
    colframe=black!35,
    boxrule=0.5pt,
    arc=1pt,
    left=5pt,
    right=5pt,
    top=2pt,
    bottom=2pt,
    boxsep=0pt,
    before skip=0pt,
    after skip=0pt
]
\noindent\textbf{RQ1 Observation.}
Diagnosis depends on the input and attack type. Source code favors sysvar-spoofing diagnosis, while transaction records favor missing-signer diagnosis but cannot reliably distinguish attacks from legitimate rotations.
\end{tcolorbox}
\vspace{-0.1 cm}

\subsection*{RQ2: Alarm Versus Diagnosis}
\label{sec:results-rq2}
\begin{figure*}[t]
  \centering
  \includegraphics[width=0.88\textwidth]
  {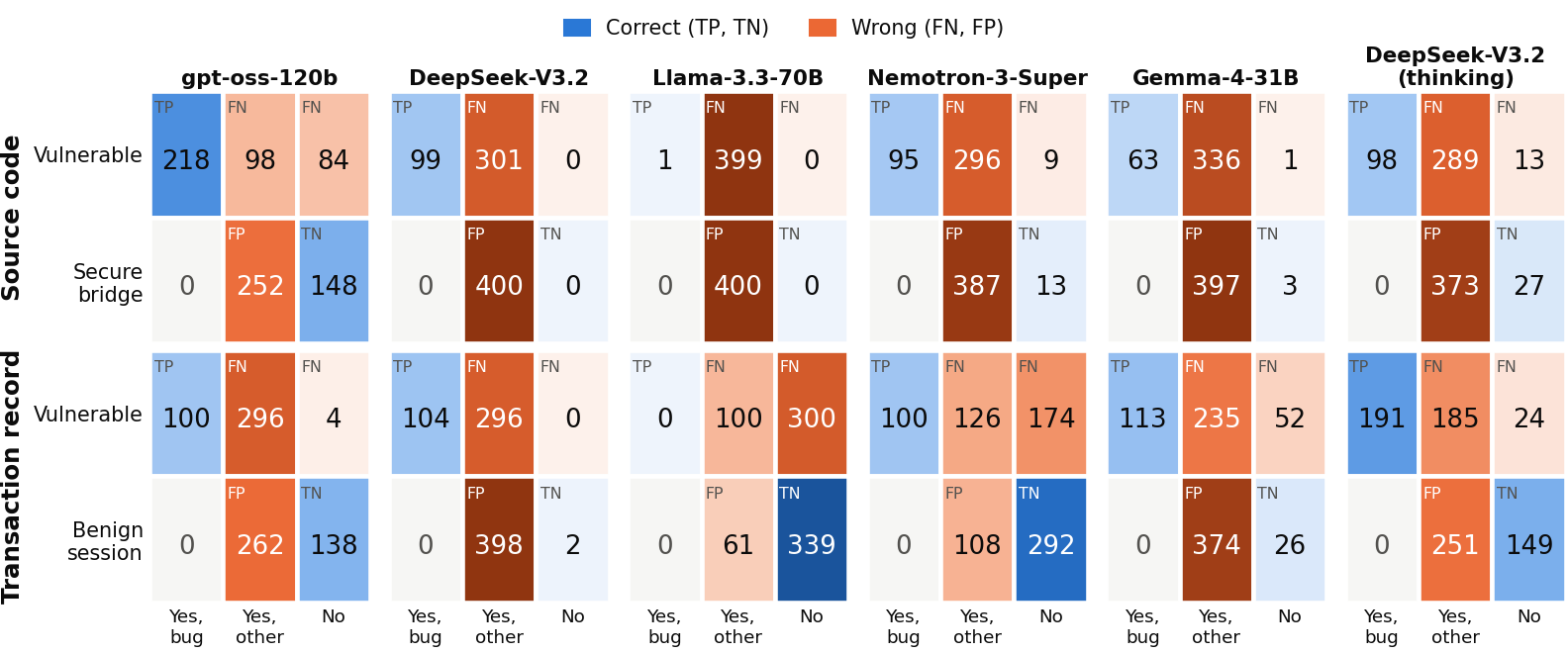}
  \vspace{-3mm}
  \caption{Response outcomes across configurations and input types, pooling all four attacks (400 vulnerable and 400 secure or benign runs per configuration). \emph{Yes, bug} indicates an alarm correctly identifying the planted vulnerability; \emph{Yes, other} indicates an alarm with an incorrect diagnosis; and \emph{No} indicates no alarm. On vulnerable inputs, only \emph{Yes, bug} counts as a diagnosed true positive. On secure or benign inputs, any affirmative response counts as a false positive. \vspace{-1mm}
  }
  \label{fig:confusion}
\end{figure*}

RQ2 examines whether models that raise alarms also identify the planted vulnerability. Figure~\ref{fig:confusion} distinguishes three response categories: \emph{Yes, bug} (correct alarm and diagnosis), \emph{Yes, other} (alarm with an incorrect diagnosis), and \emph{No} (no alarm). On vulnerable inputs, only \emph{Yes, bug} counts as a diagnosed true positive, while the other two categories count as diagnostic misses. Table~\ref{tab:detection} compares performance at both levels.
\vspace{0.15cm}

\noindent\textbf{Alarm reliability.} In Table~\ref{tab:detection}, five configurations flag 97--100\% of vulnerable source-code inputs, yet achieve only 50--52\% accuracy. In Figure~\ref{fig:confusion}, they produce 373--400 false positives (\emph{Yes, other} column) across 400 secure-bridge evaluations, corresponding to false-positive rates of 93.25--100\%. This indicates that their high vulnerability recall is accompanied by poor discrimination between vulnerable and secure implementations. Since accuracy measures correct classifications across both vulnerable and secure inputs, \textit{their near-perfect recall is offset by an equally high false-positive rate}. An always-positive detector would achieve recall of 1.00, F1 of 0.67, and accuracy of 0.50 on this balanced dataset, closely matching several configurations without distinguishing vulnerable from secure implementations. Thus, high alarm recall provides little evidence of meaningful vulnerability detection when nearly every input is classified as vulnerable. Among the six configurations, gpt-oss-120b shows better discrimination, withholding alarms on roughly one-third of secure-bridge runs while achieving the highest source-code diagnosis count (218 of 400). Nevertheless, its flagged accuracy remains at 58\%, indicating that substantial false positives persist even in the best-performing source-code configuration.
\vspace{0.15cm}

\noindent\textbf{Diagnostic correctness.} Requiring models to identify the planted vulnerability exposes a substantial gap between raising alarms and diagnosing attacks. As shown in Figure~\ref{fig:confusion}, DeepSeek-V3.2 without thinking produces 301 incorrect diagnoses (\emph{Yes, other}) against only 99 correct diagnoses (\emph{Yes, bug}) across 400 vulnerable source-code runs. The disparity is even greater for Llama-3.3-70B, with 399 incorrect diagnoses and only one correct diagnosis. Transaction records show a similar limitation: gpt-oss-120b produces 296 incorrect diagnoses against 100 correct diagnoses, while DeepSeek-V3.2 with thinking enabled achieves 191 correct diagnoses but still produces 185 incorrect ones. As reported in Table~\ref{tab:detection}, diagnosed recall consequently ranges from 0.00-0.55 for source code and 0.00-0.48 for transactions. These findings show that models frequently recognize vulnerable inputs without identifying the underlying security failure. Such responses contribute to flagged recall but count as diagnostic misses, demonstrating that a positive alarm alone provides limited evidence of correct vulnerability identification.

\vspace{2mm}

\begin{tcolorbox}[
    colback=blue!15,
    colframe=black!35,
    boxrule=0.5pt,
    arc=1pt,
    left=5pt,
    right=5pt,
    top=2pt,
    bottom=2pt,
    boxsep=0pt,
    before skip=0pt,
    after skip=0pt
]
\textbf{RQ2 Observation.} High alarm recall does not establish reliable detection: models frequently flag secure inputs or identify the wrong flaw. Correct diagnosis and discrimination from benign activity remain major limitations.
\end{tcolorbox}

\subsection*{RQ3: Effect of Code Comments}
\label{sec:results-rq3}

\begin{table}[t]
  \centering
  \normalsize
  \caption{Runs (of 400) in which each configuration named the planted bug in the source code, without
  comments and with the program's original comments (4 attacks $\times$ 100 runs).}
  \vspace{-0.2 cm}
  \label{tab:comments}
  \begin{tabular}{lrr}
\toprule
Model & No comments & Original comments \\
\midrule
gpt-oss-120b & 218 & 306 \\
DeepSeek-V3.2 & 99 & 202 \\
Llama-3.3-70B & 1 & 188 \\
Nemotron-3-Super & 95 & 214 \\
Gemma-4-31B & 63 & 269 \\
DeepSeek-V3.2 (thinking) & 98 & 254 \\
\midrule
All models (of 2,400) & 574 (24\%) & 1,433 (60\%) \\
\bottomrule
\end{tabular}
  \vspace{-0.1 cm}
\end{table}
\begin{figure*}[t]
  \centering
  \includegraphics[width=0.88\textwidth]{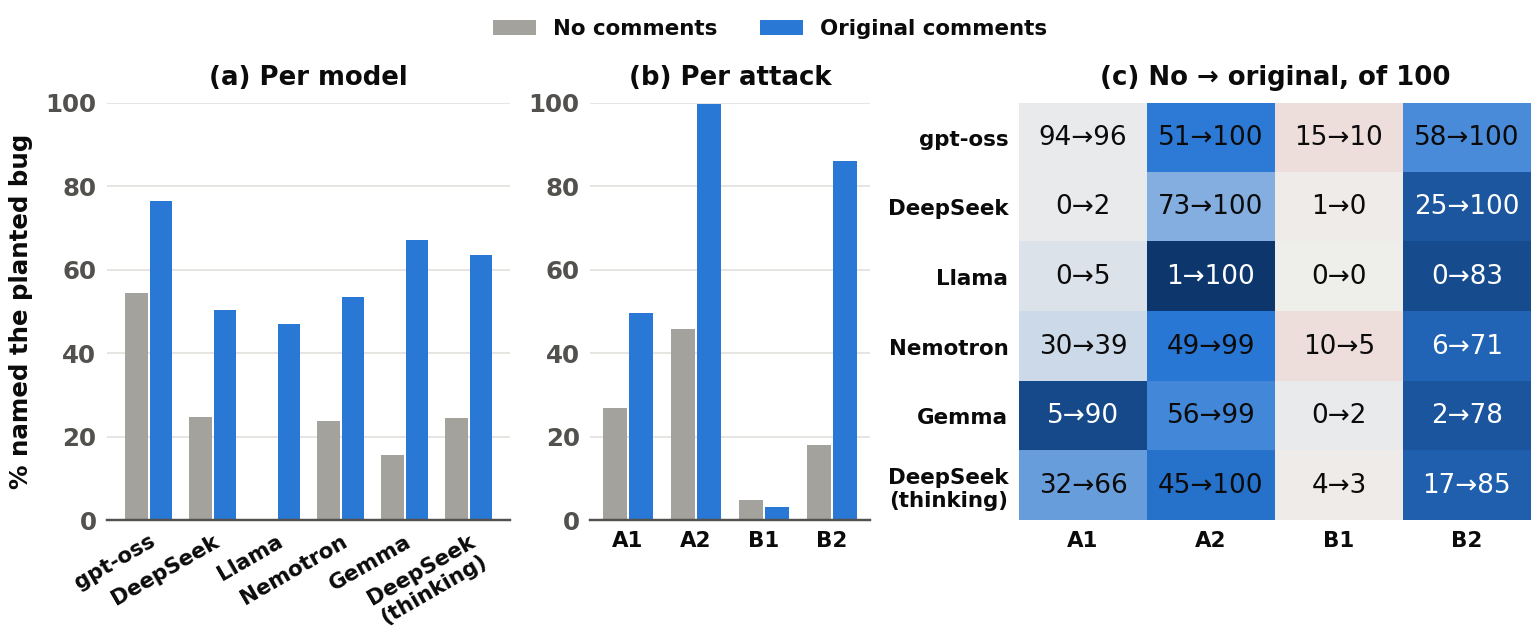} 
  \vspace{-3mm}
    \caption{Effect of code comments on vulnerability diagnosis across six configurations and four attack scenarios (100 runs per configuration and attack). Bars compare diagnosis rates with and without comments, while the heatmap shows the counts. Improvements are concentrated in A1, A2, and B2, with little or no improvement in B1. \vspace{-1.5mm}}
  \label{fig:comments}
\end{figure*}
RQ3 examines the effect of original code comments on vulnerability diagnosis, with executable logic unchanged. Figure~\ref{fig:comments} compares per-attack diagnosis counts with and without comments, while Table~\ref{tab:comments} summarizes performance across configurations.
\vspace{0.1 cm}

\noindent\textbf{Overall effect of comments.} As reported in Table~\ref{tab:comments}, restoring comments increases correct diagnoses from 574 to 1,433 of 2,400 runs (24\% to 60\%). All six configurations improve, although the gains differ considerably. The largest absolute increase occurs for Gemma-4-31B, from 63 to 269 of 400 runs, followed by Llama-3.3-70B, from only 1 to 188. Even gpt-oss-120b, the strongest configuration without comments, improves from 218 to 306. Since the executable code and evaluation conditions remain unchanged, these differences demonstrate substantial sensitivity to the accompanying comments. However, higher diagnosis counts do not necessarily establish stronger vulnerability reasoning: a model may correctly name a missing check because the comments describe it rather than because it inferred the flaw from executable logic.
\vspace{0.1 cm}

\noindent\textbf{Attack-specific effects.} As shown in Figure~\ref{fig:comments}, the improvement is concentrated in scenarios where comments provide clues about the planted vulnerability. For A2 (sysvar spoofing) and B2 (arbitrary CPI), comments at the vulnerable functions explicitly describe the missing validation checks, and all configurations achieve 99-100 and 71-100 correct diagnoses of 100 runs, respectively. In contrast, B1 (missing signer) has no comment identifying the omitted authorization check and shows no improvement, with diagnoses decreasing from 30 to 20 of 600 runs across configurations. A1 (mint confusion) exhibits substantial but less uniform gains: Gemma-4-31B improves from 5 to 90 of 100 runs, while DeepSeek-V3.2 with thinking enabled rises from 32 to 66. Unlike A2 and B2, comments elsewhere in the program point to the mint without directly identifying the missing check. These differences suggest that both explicit descriptions of vulnerabilities and indirect contextual cues can influence model diagnoses. Crucially, the absence of improvement on B1 contrasts with the near-perfect A2 results, indicating that the overall gain depends strongly on what the comments reveal rather than reflecting a uniform improvement in source-code analysis.

\vspace{2mm}

\begin{tcolorbox}[
    colback=blue!15,
    colframe=black!35,
    boxrule=0.5pt,
    arc=1pt,
    left=5pt,
    right=5pt,
    top=2pt,
    bottom=2pt,
    boxsep=0pt,
    before skip=0pt,
    after skip=0pt
]
\textbf{RQ3 Observation.} Comments raise correct diagnoses from 24\% to 60\%, but improvements concentrate where comments reveal the missing checks. Higher diagnosis rates therefore do not necessarily indicate better analysis of executable code.
\end{tcolorbox}

\section{Conclusion and Future Work}
\label{sec:conclusion}
\vspace{0.15 cm}
We evaluated six open-weight LLM configurations on four exploitable Solana bridge variants, collecting 16,800 responses across source-code and transaction-based evaluations. The results reveal a substantial gap between alarms and correct diagnoses: five configurations flag nearly every secure bridge input, while only 24\% of vulnerable source-code responses identify the planted flaw. Transaction-based diagnosis varies by attack type, with missing-signer exploits frequently identified but difficult to distinguish from legitimate administrator changes. Restoring original comments increases source-code diagnosis to 60\%, largely where comments reveal missing validation checks. These findings highlight the need to evaluate diagnostic correctness separately from alarms and control information disclosed in model inputs. LLM-generated findings therefore require independent verification before informing security decisions. Future work should investigate the effects of misleading, outdated, and adversarial comments, including fabricated vulnerabilities and false security assurances, to assess model reliance on textual cues rather than executable logic. Further studies should examine deployed bridges with interacting vulnerabilities, richer transaction metadata, and comparisons with static analysis tools. Human-reviewed benchmarks should also distinguish alternative valid findings from errors.

\section*{GenAI Usage Disclosure}
Generative AI tools were used to support various stages of this research, including brainstorming research ideas, identifying relevant literature, assisting with code development and debugging, and refining manuscript content and presentation. Large language models were also incorporated into the experimental methodology, including the independent grading of model-generated vulnerability diagnoses, as detailed in the evaluation section. AI-assisted outputs were reviewed and verified by the authors, and all experimental decisions, interpretation of results, and final conclusions remain the responsibility of the authors.

\vspace{-2mm}

\bibliographystyle{ACM-Reference-Format}
\bibliography{references}

@inproceedings{gritti2023confusum,
  author    = {Fabio Gritti and Nicola Ruaro and Robert McLaughlin and Priyanka Bose and Dipanjan Das and Ilya Grishchenko and Christopher Kruegel and Giovanni Vigna},
  title     = {Confusum Contractum: Confused Deputy Vulnerabilities in {Ethereum} Smart Contracts},
  booktitle = {32nd USENIX Security Symposium (USENIX Security 23)},
  year      = {2023},
  month     = aug,
  pages     = {1793--1810},
  publisher = {USENIX Association},
  url       = {https://www.usenix.org/conference/usenixsecurity23/presentation/gritti}
}

@inproceedings{xie2022zkbridge,
  author    = {Tiancheng Xie and Jiaheng Zhang and Zerui Cheng and Fan Zhang and Yupeng Zhang and Yongzheng Jia and Dan Boneh and Dawn Song},
  title     = {{zkBridge}: Trustless Cross-chain Bridges Made Practical},
  booktitle = {Proceedings of the 2022 ACM SIGSAC Conference on Computer and Communications Security},
  series    = {CCS '22},
  year      = {2022},
  pages     = {3003--3017},
  publisher = {Association for Computing Machinery},
  doi       = {10.1145/3548606.3560652},
  url       = {https://doi.org/10.1145/3548606.3560652}
}

@inproceedings{huang2026xrbc,
  author    = {Yue Huang and Xin Wang and Haibin Zhang and Sisi Duan},
  title     = {Cross-Consensus Reliable Broadcast and its Applications},
  booktitle = {Proceedings of the Network and Distributed System Security Symposium (NDSS)},
  year      = {2026},
  month     = feb,
  publisher = {Internet Society},
  doi       = {10.14722/ndss.2026.230207},
  url       = {https://doi.org/10.14722/ndss.2026.230207}
}

@inproceedings{wei2022chainofthought,
  author    = {Jason Wei and Xuezhi Wang and Dale Schuurmans and Maarten Bosma and Brian Ichter and Fei Xia and Ed H. Chi and Quoc V. Le and Denny Zhou},
  title     = {Chain-of-Thought Prompting Elicits Reasoning in Large Language Models},
  booktitle = {Advances in Neural Information Processing Systems},
  volume    = {35},
  year      = {2022},
  pages     = {24824--24837},
  publisher = {Curran Associates, Inc.},
  url       = {https://proceedings.neurips.cc/paper_files/paper/2022/hash/9d5609613524ecf4f15af0f7b31abca4-Abstract-Conference.html}
}

@inproceedings{turpin2023unfaithful,
  author    = {Miles Turpin and Julian Michael and Ethan Perez and Samuel R. Bowman},
  title     = {Language Models Don't Always Say What They Think: Unfaithful Explanations in Chain-of-Thought Prompting},
  booktitle = {Advances in Neural Information Processing Systems},
  volume    = {36},
  year      = {2023},
  pages     = {74952--74965},
  publisher = {Curran Associates, Inc.},
  url       = {https://proceedings.neurips.cc/paper_files/paper/2023/hash/ed3fea9033a80fea1376299fa7863f4a-Abstract-Conference.html}
}

@misc{chainalysis2022bridges,
  author       = {{Chainalysis Team}},
  title        = {Vulnerabilities in Cross-chain Bridge Protocols Emerge as Top Security Risk},
  howpublished = {Chainalysis Blog},
  year         = {2022},
  month        = aug,
  url          = {https://blog.chainalysis.com/reports/cross-chain-bridge-hacks-2022/}
}

@misc{halborn2022wormhole,
  author       = {Behnke, Rob},
  title        = {Explained: The {Wormhole} Hack (February 2022)},
  howpublished = {Halborn Blog},
  year         = {2022},
  month        = feb,
  day          = {4},
  url          = {https://halborn.com/explained-the-wormhole-hack-february-2022}
}

@misc{certik2022wormhole,
  author       = {{CertiK}},
  title        = {Wormhole Bridge Exploit Incident Analysis},
  year         = {2022},
  howpublished = {Blog post},
  url          = {https://www.certik.com/resources/blog/wormhole-bridge-exploit-incident-analysis},
  lastaccessed = {November 11, 2026}
}

@misc{certik2022cashio,
  author       = {{CertiK}},
  title        = {Cashio App Incident Analysis},
  year         = {2022},
  howpublished = {Blog post},
  url          = {https://www.certik.com/blog/cashio-app-incident-analysis},
  lastaccessed = {November 11, 2026}
}

@misc{solana2020,
  author       = {{Solana Foundation}},
  year         = {2020},
  url          = {https://solana.com/docs/core/pda}
}

@misc{neodyme2022anker,
  author       = {{Neodyme}},
  title        = {Security Audit -- {Anker} on {Solana}},
  year         = {2022},
  month        = apr,
  day          = {6},
  url          = {https://www.neodyme.io/reports/Anker.pdf}
}

@inproceedings{augusto2024interoperability,
  author    = {Andr{\'e} Augusto and Rafael Belchior and Miguel Correia and Andr{\'e} Vasconcelos and Luyao Zhang and Thomas Hardjono},
  title     = {{SoK}: Security and Privacy of Blockchain Interoperability},
  booktitle = {2024 IEEE Symposium on Security and Privacy (SP)},
  year      = {2024},
  pages     = {3840--3865},
  publisher = {IEEE},
  doi       = {10.1109/SP54263.2024.00255},
  url       = {https://doi.org/10.1109/SP54263.2024.00255}
}

@inproceedings{sun2024tokens,
  author    = {Tianle Sun and Ningyu He and Jiang Xiao and Yinliang Yue and Xiapu Luo and Haoyu Wang},
  title     = {All Your Tokens are Belong to Us: Demystifying Address Verification Vulnerabilities in {Solidity} Smart Contracts},
  booktitle = {33rd USENIX Security Symposium (USENIX Security 24)},
  year      = {2024},
  month     = aug,
  pages     = {3567--3584},
  publisher = {USENIX Association},
  isbn      = {978-1-939133-44-1},
  url       = {https://www.usenix.org/conference/usenixsecurity24/presentation/sun-tianle}
}

@inproceedings{andreina2024defying,
  author    = {S{\'e}bastien Andreina and Tobias Cloosters and Lucas Davi and Jens-Rene Giesen and Marco Gutfleisch and Ghassan Karame and Alena Naiakshina and Houda Naji},
  title     = {Defying the Odds: {Solana}'s Unexpected Resilience in Spite of the Security Challenges Faced by Developers},
  booktitle = {Proceedings of the 2024 on ACM SIGSAC Conference on Computer and Communications Security},
  series    = {CCS '24},
  year      = {2024},
  pages     = {4226--4240},
  publisher = {Association for Computing Machinery},
  doi       = {10.1145/3658644.3670333},
  url       = {https://doi.org/10.1145/3658644.3670333}
}

@inproceedings{cui2022vrust,
  author    = {Siwei Cui and Gang Zhao and Yifei Gao and Tien Tavu and Jeff Huang},
  title     = {{VRust}: Automated Vulnerability Detection for {Solana} Smart Contracts},
  booktitle = {Proceedings of the 2022 ACM SIGSAC Conference on Computer and Communications Security},
  series    = {CCS '22},
  year      = {2022},
  pages     = {639--652},
  publisher = {Association for Computing Machinery},
  doi       = {10.1145/3548606.3560552},
  url       = {https://doi.org/10.1145/3548606.3560552}
}

@inproceedings{smolka2023fuzzdelsol,
  author    = {Sven Smolka and Jens-Rene Giesen and Pascal Winkler and Oussama Draissi and Lucas Davi and Ghassan Karame and Klaus Pohl},
  title     = {Fuzz on the Beach: Fuzzing {Solana} Smart Contracts},
  booktitle = {Proceedings of the 2023 ACM SIGSAC Conference on Computer and Communications Security},
  series    = {CCS '23},
  year      = {2023},
  pages     = {1197--1211},
  publisher = {Association for Computing Machinery},
  doi       = {10.1145/3576915.3623178},
  url       = {https://doi.org/10.1145/3576915.3623178}
}

@inproceedings{wang2024smartinv,
  author    = {Sally Junsong Wang and Kexin Pei and Junfeng Yang},
  title     = {{SmartInv}: Multimodal Learning for Smart Contract Invariant Inference},
  booktitle = {2024 IEEE Symposium on Security and Privacy (SP)},
  year      = {2024},
  pages     = {2217--2235},
  publisher = {IEEE},
  doi       = {10.1109/SP54263.2024.00126},
  url       = {https://doi.org/10.1109/SP54263.2024.00126}
}

@inproceedings{liu2025propertygpt,
  author    = {Ye Liu and Yue Xue and Daoyuan Wu and Yuqiang Sun and Yi Li and Miaolei Shi and Yang Liu},
  title     = {{PropertyGPT}: {LLM}-driven Formal Verification of Smart Contracts through Retrieval-Augmented Property Generation},
  booktitle = {Proceedings of the Network and Distributed System Security Symposium (NDSS)},
  year      = {2025},
  month     = feb,
  publisher = {Internet Society},
  doi       = {10.14722/ndss.2025.241357},
  url       = {https://doi.org/10.14722/ndss.2025.241357}
}

@article{huang2026paravul,
  author    = {Tenghui Huang and Jinbo Wen and Jiawen Kang and Siyong Chen and Zhengtao Li and Tao Zhang and Dongning Liu and Jiacheng Wang and Chengjun Cai and Yinqiu Liu},
  title     = {{ParaVul}: A Parallel Large Language Model and Retrieval-Augmented Framework for Smart Contract Vulnerability Detection},
  journal   = {IEEE Transactions on Information Forensics and Security},
  year      = {2026},
  volume    = {21},
  pages     = {5017--5030},
  doi       = {10.1109/TIFS.2026.3694661},
  url       = {https://doi.org/10.1109/TIFS.2026.3694661}
}

@article{wei2026hktsmartaudit,
  author    = {Zhiyuan Wei and Jing Sun and Zijian Zhang and Zhe Hou and Xianhao Zhang and Meng Li and Yuqiang Sun and Daoyuan Wu and Yang Liu and Chunmiao Li and Mingchao Wan and Jin Dong},
  title     = {{HKT-SmartAudit}: Distilling Lightweight Models for Smart Contract Auditing},
  journal   = {IEEE Transactions on Information Forensics and Security},
  year      = {2026},
  volume    = {21},
  pages     = {4446--4459},
  doi       = {10.1109/TIFS.2026.3683274},
  url       = {https://doi.org/10.1109/TIFS.2026.3683274}
}

@article{chen2025smartguard,
  author    = {Jie Chen and Liangmin Wang and Huijuan Zhu},
  title     = {{SmartGuard}: Making Prediction Verifiable Through Transaction Sequences for Smart Contract Vulnerability Detection},
  journal   = {IEEE Transactions on Information Forensics and Security},
  year      = {2025},
  volume    = {20},
  pages     = {6117--6132},
  doi       = {10.1109/TIFS.2025.3580224},
  url       = {https://doi.org/10.1109/TIFS.2025.3580224}
}

@inproceedings{zhang2020txspector,
  author    = {Mengya Zhang and Xiaokuan Zhang and Yinqian Zhang and Zhiqiang Lin},
  title     = {{TXSPECTOR}: Uncovering Attacks in {Ethereum} from Transactions},
  booktitle = {29th USENIX Security Symposium (USENIX Security 20)},
  year      = {2020},
  month     = aug,
  pages     = {2775--2792},
  publisher = {USENIX Association},
  url       = {https://www.usenix.org/conference/usenixsecurity20/presentation/zhang-mengya}
}

@misc{solana_bridge_llm_artifacts,
  author       = {Rahman, Yasir and Samiha, Ishraq and Sakib, Shahnewaz Karim and Das, Anindya Bijoy},
  title        = {LLMs as Validators for Cross-Chain Bridges: Evaluating Source-Level and Transaction-Level Attack Detection on Solana},
  howpublished = {\url{https://github.com/ChyYasir/solana-bridge-llm-detection}},
  note         = {Artifact repository},
  year         = {2026}
}

@article{openai2025gptoss,
  title={gpt-oss-120b \& gpt-oss-20b model card},
  author={Agarwal, Sandhini and Ahmad, Lama and Ai, Jason and Altman, Sam and Applebaum, Andy and Arbus, Edwin and Arora, Rahul K and Bai, Yu and Baker, Bowen and Bao, Haiming and others},
  journal={arXiv preprint arXiv:2508.10925},
  year={2025}
}

@article{nvidia2026nemotron,
  title={Nemotron 3 super: Open, efficient mixture-of-experts hybrid mamba-transformer model for agentic reasoning},
  author={Chandiramani, Aakshita and Blakeman, Aaron and Olaoye, Abdullahi and Gupta, Abhibha and Somasamudramath, Abhilash and Khattar, Abhinav and Adesoba, Adeola and Renduchintala, Adi and Asif, Adil and Agrawal, Aditya and others},
  journal={arXiv preprint arXiv:2604.12374},
  year={2026}
}

@misc{meta2024llama33,
  author       = {{Meta}},
  title        = {{Llama 3.3 Model Card}},
  year         = {2024},
  url          = {https://github.com/meta-llama/llama-models/blob/main/models/llama3_3/MODEL_CARD.md},
  note         = {Official model documentation}
}

@article{deepseek2025v32,
  title={Deepseek-v3. 2: Pushing the frontier of open large language models},
  author={Liu, Aixin and Mei, Aoxue and Lin, Bangcai and Xue, Bing and Wang, Bingxuan and Xu, Bingzheng and Wu, Bochao and Zhang, Bowei and Lin, Chaofan and Dong, Chen and others},
  journal={arXiv preprint arXiv:2512.02556},
  year={2025}
}

@article{gemma2026gemma4,
  title={Gemma 4 technical report},
  author={Team, Gemma and Abd, Sherif El and Aggarwal, Vaibhav and Algayres, Robin and Andreev, Alek and Bachem, Olivier and Ballantyne, Ian and Brick, Cormac and C{\u{a}}rbune, Victor and Casbon, Michelle and others},
  journal={arXiv preprint arXiv:2607.02770},
  year={2026}
}

@misc{deepinfra2026,
  author       = {{DeepInfra}},
  title        = {{DeepInfra: AI Inference Platform}},
  year         = {2026},
  howpublished = {\url{https://deepinfra.com/}},
  note         = {Accessed: 2026-11-15}
}

@misc{anthropic2026opus55,
  author       = {{Anthropic}},
  title        = {{Claude Opus 5.5}},
  year         = {2026},
  howpublished = {\url{https://platform.claude.com/docs/en/models/opus-5-5/overview}},
  note         = {Official model documentation}
}

@misc{google2026gemini38flash,
  author       = {{Google DeepMind}},
  title        = {{Gemini 3.8 Flash Model Card}},
  year         = {2026},
  month        = sep,
  howpublished = {\url{https://deepmind.google/models/model-cards/gemini-3-8-flash/}},
  note         = {Official model card}
}

@inproceedings{zhang2024security,
  title={Security of cross-chain bridges: Attack surfaces, defenses, and open problems},
  author={Zhang, Mengya and Zhang, Xiaokuan and Zhang, Yinqian and Lin, Zhiqiang},
  booktitle={Proceedings of the 27th International Symposium on Research in Attacks, Intrusions and Defenses},
  pages={298--316},
  year={2024}
}

@inproceedings{boi2025prompt,
  title={Prompt Engineering vs. Fine-Tuning for LLM-Based Vulnerability Detection in Solana and Algorand Smart Contracts},
  author={Boi, Biagio and Esposito, Christian},
  booktitle={2025 7th International Conference on Blockchain Computing and Applications (BCCA)},
  pages={391--398},
  year={2025},
  organization={IEEE}
}

@inproceedings{hu2023large,
  title={Large language model-powered smart contract vulnerability detection: New perspectives},
  author={Hu, Sihao and Huang, Tiansheng and {\.I}lhan, Fatih and Tekin, Selim Furkan and Liu, Ling},
  booktitle={2023 5th IEEE International Conference on Trust, Privacy and Security in Intelligent Systems and Applications (TPS-ISA)},
  pages={297--306},
  year={2023},
  organization={IEEE}
}

@article{xu2026large,
  title={Large language models for cyber security: A systematic literature review},
  author={Xu, HanXiang and Wang, ShenAo and Li, Ningke and Wang, Kailong and Zhao, Yanjie and Chen, Kai and Yu, Ting and Liu, Yang and Wang, HaoYu},
  journal={ACM Transactions on Software Engineering and Methodology},
  volume={35},
  number={11},
  pages={1--61},
  year={2026},
  publisher={ACM, New York, NY}
}

\end{document}